\documentclass[aps,prb,twocolumn,floatfix,amsmath,amssymb,superscriptaddress,showpacs]{revtex4}
\usepackage{times}
\usepackage{standalone}
\usepackage{graphicx}
\usepackage{dcolumn}
\usepackage{bm}
\usepackage{color}
\usepackage{float}
\usepackage{mathrsfs}
\usepackage[breaklinks]{hyperref}

\def\lB{\left[}\def\rB{\right]}
\def\lP{\left(}\def\rP{\right)}
\def\lC{\left\{}
\def\lA{\left|}\def\rA{\right|}
\def\rN{\right .}
\def\lV{\left\langle}\def\rV{\right\rangle}
\def\bra#1{\lV #1 \rA}
\def\ket#1{\lA #1 \rV}
\def\braket#1#2{\lV #1 \middle| #2 \rV}
\def\V#1{\bm{#1}}
\def\Vhat#1{\hat{\bm{#1}}}
\def\mtrx{\lB\begin{array}{rrrrrrrrrrrrrr}}
\def\emtrx{\end{array}\rB}
\def\sZ{\mathbb{Z}}
\def\sR{\mathbb{R}}
\def\pad#1{\quad \text{#1}\quad }
\def\op#1{\mathscr{#1}}

\begin{document}

\title{Phases in two dimensional $p_x+ip_y$ superconducting systems with next-nearest-neighbor interactions}
\author{Antonio Russo}
\affiliation{Department of Physics and Astronomy, University of
California Los Angeles, Los Angeles, California 90095-1547, USA}
\author{Sudip Chakravarty}
\affiliation{Department of Physics and Astronomy, University of
California Los Angeles, Los Angeles, California 90095-1547, USA}
\date{November 8, 2013}

\begin{abstract}
A chiral $p_x+ip_y$ superconductor on a square lattice with nearest and next-nearest hopping and pairing terms is considered.
Gap closures, as various parameters of the system are varied, are found analytically and used to identify the topological phases.
The phases are characterized by Chern numbers (ranging from $-3$ to $3$),
and (numerically) by response to introduction of weak disorder, edges,
and magnetic fields in an extreme type-II limit, focusing on the 
low-energy modes (which presumably become zero-energy Majorana modes for large lattices and separations).
Several phases are found, including a phase with Chern number $3$ that cannot be thought of in terms of a single range of
interaction, and phase with Chern number $2$ that may host an additional, disorder resistant, Majorana mode. The energies of the vortex
quasiparticle modes were found to oscillate as vortex position varied.  The spatial length scale of these
oscillations was found for various points in the Chern number $3$ phase which increased as criticality was approached. 
\end{abstract}
\maketitle
\section{Introduction}
Recently, there has been much interest in topological features of various condensed
matter systems, in particular Majorana fermions.
\cite{kouwenhoven-2012,kitaev-2001,wilczek-2009,moore-read-1991,fu-kane-2008,sau-lutchyn-2010,sau-tewari-2010,g_Refael_von_Oppen_Fisher_2011,alicea-2010,lutchyn-sau-2010,degottardi-diptiman-2011,Alicea_2012,read-green-2000,ivanov-2001,roy-2010}
Majorana fermions satisfy $\gamma^\dag=\gamma$; that is, they are their own antiparticle.
In systems with particle-hole symmetry, their energy is therefore pinned to zero. Consequently,
Majoranas can only be destroyed by pairing with another and hybridizing into a Dirac fermion. \par
We focus on chiral $p_x+ip_y$ superconductors. 
In continuum models with nonzero Chern numbers, zero-energy Majoranas
develop around defects, such as vortices.\cite{kopnin-salomaa-1991,Alicea_2012,read-green-2000,ivanov-2001,volovik-1999,caroli-1964,hasan-kane-2010,roy-2010}
When the vortices are well-separated, the associated Majoranas
are protected from local perturbations, which could be useful in
quantum computers. Majoranas are also expected in lattice models of chiral superconductors;
if the gap is of the form $\sin(nk_x)+i\sin(nk_y)$, where $n$ is the range of the interaction,
it reduces to $n\partial_x+in\partial_y$ in the continuum limit.\par
\begin{figure}
\includegraphics{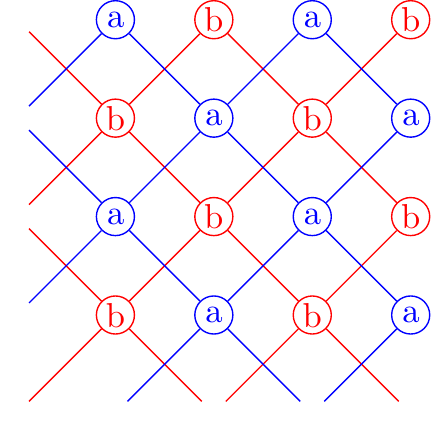}
\caption{(color online) When only next-nearest neighbor interactions are present in a 2 dimensional lattice model,
two sublattices form, each independently responding to defects, giving pairs of defect modes. When nearest-neighbor
interactions are turned on, the pairs of defect modes persist.
\label{2nn-lattice}}
\end{figure}

In contrast to the continuum case where the range of the interaction, $n$, simply rescales the gap 
function, the range plays a more interesting role on the lattice. Previous work\cite{niu-chakravarty-2012}
suggested that the inclusion of longer-ranged interactions leads to novel phases. These longer-ranged interactions
in general give rise to larger Chern numbers in a way that can be most easily understood when
all interactions are of the same range. When all interactions are of the same range, a number of
noninteracting sublattices, $S(n)$, form. For example, in FIG.~\ref{2nn-lattice}, two sublattices
form when only next-nearest neighbor terms are present. As separate systems, each sublattice has
its own Chern number, either $0$ or $1$. Therefore, the Chern number for the whole system is either $0$ or $S(n)$.\par
\begin{figure}[t]
\includegraphics{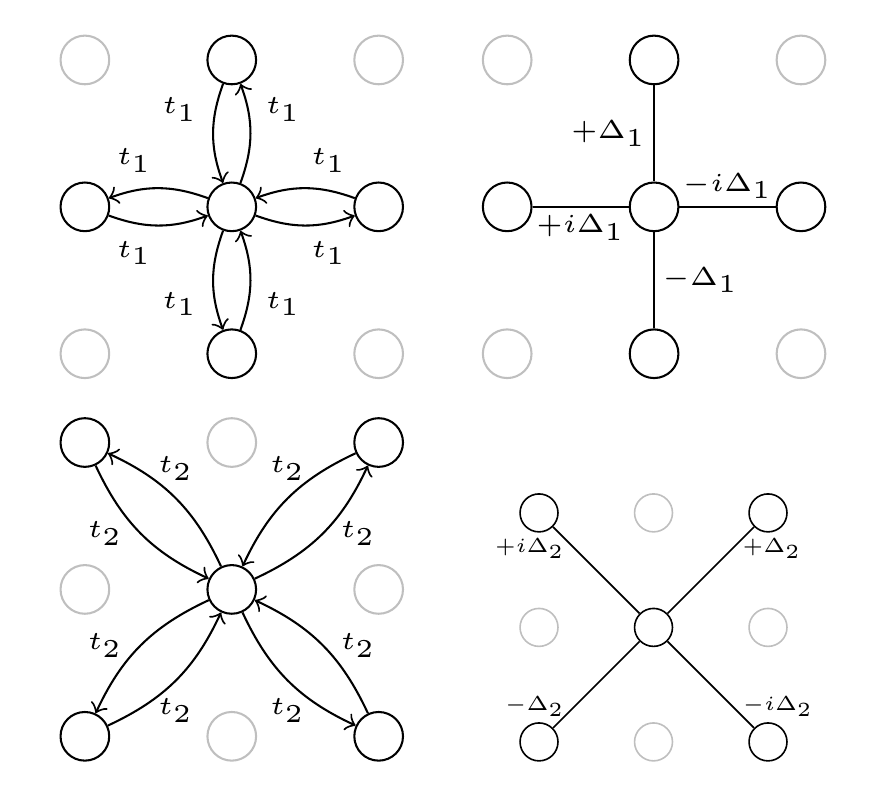}
\caption{Visualization of nearest- (and second nearest-) neighbor hopping ($t$) and pairing ($\Delta$) terms. 
\label{hop-pair-illus}}
\end{figure}

To explore the more complicated case of interactions of different ranges,  we study a square
lattice with a combination of nearest-neighbor (NN) and next-nearest-neighbor (NNN) hopping
(respectively, $t_1$ and $t_2$), and $p_x+ip_y$ pairing (respectively, $\Delta_1$ and $\Delta_2$)
terms. The system is kept at chemical potential $\mu$. Both the hopping and pairing terms are illustrated in FIG.~\ref{hop-pair-illus}.
\par
The five parameters $t_1$, $t_2$, $\Delta_1$, $\Delta_2$, and $\mu$ constitute a parameter space
rich enough to include the well-known BEC and BCS superconducting systems, as well as their two-sublattice
versions (i.e., purely NNN interactions). Because the BEC-BCS transition is topological in nature,
we search for the surfaces in parameter space where the bulk band gap collapses and topological phase transitions
occur.
\par

Our analysis of the system shows that the phase diagram depends only
on \emph{three} ratios of parameters: $\alpha=\frac{\Delta_1}{\Delta_2}$ and the scaled hopping
terms $t_1/\mu$, $t_2/\mu$. For fixed values of $\alpha$, all phase transitions are
lines in the $t_1$-$t_2$ plane. There are four such lines, three of which are independent of
$\alpha$. When a system is tuned to one of these phase transition lines, the gap in the system collapses
because a zero of $\Delta(\V k)$ is crossing the Fermi surface. The lines
constitute the phase diagram for a given value of $\alpha$, as shown in
FIG.~\ref{phase-diagram}. When $|\alpha|^2\geq 2$, the phase transition lines remain fixed,
and $\Delta(\V k)$ only has zeros at the four high-symmetry points $\Gamma$,
$X$, $Y$, and $M$ because the NNN pairing terms are not strong enough to introduce
any zeros into $\Delta(\V k)$. The topology of the system is unaffected by the weak NNN
pairing. When $|\alpha|^2<2$, four additional zeros are introduced into $\Delta(\V k)$, permitting
larger Chern numbers. An analytical calculation finds all Chern numbers possible with NN and NNN terms;
they range from $-3$ to $+3$. Chern number $\pm 4$, while conceivably possible with NNN pairing terms,
cannot be obtained with just NNN hopping terms for the same reason that Chern number $2$ cannot
be obtained with just NN pairing and hopping terms. However, the system does take on Chern number $3$,
which is surprising because purely NNN interactions yield Chern number $2$. 
\par
The numerical aspect of the present work characterizes the response of the model system
to defects in different phases. In particular, we focus on
characterizing the low-energy response, i.e., identifying the the fundamental excitations
of the system. For the numerics, we include three kinds of position-dependent terms into the Hamiltonian:
edges, on-site disorder, and magnetic fields in an extreme type-II limit with vortices in the superconducting order parameter.
Edges are introduced by adding terms of the form $O_i c^\dag_ic_i$, where $O_i$ is very large past the edge,
confining the states in the low energy spectrum. On-site disorder is added in a similar manner:
$O_i$ takes on a value of $E_d$ with probability $\frac{p}{2}$ and $-E_d$ with the
same probability, and $0$ otherwise. For the magnetic field, we assume a very long magnetic screening length so that
the magnetic field is constant, consistent with the sample being two dimensional. However, the superconducting coherence
length $\xi$ is finite, and vortices appear in the superconducting order parameter.
\par
The output of numerical simulations are the energies and wavefunctions of the quasiparticles of the Hamiltonian.
The edge modes and vortex core modes are perfectly distinct in the ideal limit of infinite separation. In the realistic
case of finite separation, the modes hybridize.  The vortex core modes interact with each other in a similar way.
The energies of the lowest vortex core modes exhibit exponentially damped oscillations as the vortices are
separated, an effect theoretically predicted\cite{Lutchyn_Galitski_Das_Sarma_2009} and numerically observed\cite{mizushima-machida-2010}
in related systems. The edges hybridize with vortices over longer length scale than the and vortices hybridize with each other.
\par
The hybridization effects also depend on the bulk parameters of the system, i.e., $t_1$, $t_2$, $\Delta_1$, $\Delta_2$, and $\mu$.
In particular, as these parameters are tuned to the phase transitions, the edge-vortex length scale diverges.
Such tuning is explored in a system with $(\Delta_1,\Delta_2)=(0.5,1.0)$ and $t_1=-2$ (energy is given in terms of the
NNN hopping strength, $t_2$) by adjusting the chemical potential $\mu$, i.e.
by moving along the path shown in FIG.~\ref{phase-diagram}, which crosses several phase transitions.
While in the Chern number $3$ portion of the phase diagram, we find that
the spatial period of vortex-vortex oscillation increase linearly with the chemical potential: $\Lambda\sim 0.8\mu+\text{constant}$.
When $\mu$ takes on values putting the system too close to the phase transition, edge-vortex hybridization destroys the
vortex-vortex oscillatory behavior. 
\par
Another issue addressed in the numerical simulation is the number of low-energy modes created around defects.
When only NNN interactions are present, one Majorana mode per vortex per sublattice forms.
When the NN terms are turned on, the Majoranas may hybridize down to zero or one
residual zero-energy mode, for even and odd Chern number, respectively\cite{roy-2010}. Interestingly, there
is some degree of protection of the additional defect mode for the Chern number $2$ phase.
The numerical simulation reveals two, apparently disorder resistant, zero energy, vortex core modes.
The Chern number $3$ phase, however, enjoys no such additional modes: only one low-energy vortex-core mode is observed in the numerical simulations.
\par
Having introduced the primary results of the paper, the remainder of the paper explains
details of our approach. First, in section \ref{analytical}, we describe the Hamiltonian used (including both
nearest and next-nearest neighbor terms) and calculate its
Chern number analytically. Next, in section \ref{defects}, we describe the defects added 
to our model Hamiltonian: edges, disorder, and magnetic fields. Finally, in
section \ref{results}, we discuss the numerical attack on the system, with defects present, and the resulting conclusions.
Additional details are in the appendices: 
a brief review of the calculation of Chern numbers (Appendix \ref{chern-review});
and a discussion of a spatial inversion symmetry of the system helpful in distinguishing different modes (Appendix \ref{spatial-inversion}).
\section{Notation and Defect-Free Analysis\label{analytical}}
\subsection{System Definition}
Here, we describe the model system: a single-band, two-dimensional, tight-binding, spinless fermion
square lattice model with mean-field superconducting order parameter $\Delta$, and Hamiltonian
\begin{multline}
H = \sum_{ij} h_{ij} c_i^\dag c_j +\frac{1}{2} \sum_{ij} \Delta_{ij} c^\dag_i c^\dag_j
+\text{h.c.} \\
 = \frac{1}{2}\sum_{ij} \mtrx c_i^\dag & c_i \emtrx \mtrx h_{ij} & \Delta_{ij} \\ \bar\Delta_{ji} & -\bar h_{ij} \emtrx \mtrx c_j\\ c_j^\dag \emtrx \\
 = \frac{1}{2}\sum_{ij} \mtrx c_i^\dag & c_i \emtrx \mathcal{H}_{ij} \mtrx c_j\\ c_j^\dag \emtrx 
 = E_g + \sum_n E_n \Psi_n^\dag \Psi_n\label{hamiltonian}
\end{multline}
 where the indices $i$ and $j$ run over all lattice sites. The lattice separation $a$ is
set to unity. The Bogoliubov-de Gennes Hamiltonian is diagonalized in the last step 
in terms of the ground state energy $E_g$, quasiparticle energies $E_n$ and operators
\begin{equation}
 \Psi_n=\sum_i \lB u^{(n)}_i c_i+v^{(n)}_i c_i^\dag \rB
\end{equation}
\indent The hopping and pairing terms are stated here explicitly and
illustrated in FIG.~\ref{hop-pair-illus}. The lattice
separation is set to $1$, and $j$ runs over all lattice sites.
\begin{align}
 h_{j,j}&= O_j-\mu \\
 h_{j,j\pm \Vhat x} = h_{j,j\pm \Vhat y} &= t_1\\
 h_{j,j\pm (\Vhat x+\Vhat y)} = h_{j,j\pm (\Vhat x-\Vhat y)} &= t_2
\end{align}
and
\begin{align}
\Delta_{j,j\pm \Vhat x} &= \pm i\Delta_1 \pad{and}&
 \Delta_{j,j\pm \Vhat y} &= \pm\Delta_1\\
h_{j,j\pm (\Vhat x-\Vhat y)} &= \pm i\Delta_2\pad{and}&
h_{j,j\pm (\Vhat x+\Vhat y)} &= \pm \Delta_2
\end{align}
with all other terms zero. The on-site term is separated into the chemical potential $\mu$, and all other on-site
terms $O_j$, such as disorder and edges.
\par
For both the NN and NNN pairing terms, the phase of
the order parameter advances under counter-clockwise rotation, creating the
chirality of the order parameter.\footnote{
The case of opposite chirality, in which the NNN pairing has chirality opposite of the NN pairing, was investigated analytically.
It is not included because it is not clear if such terms are even physical. The results are surprisingly similar to the same chirality
case: Chern numbers range from $+3$ to $-3$.} The pairing terms explicitly break time-reversal
symmetry, putting the two dimensional system in Altland-Zirnbauer\cite{altland-zirnbauer-1997}
symmetry class D, with topological classification $\sZ$ given by
the Chern number.\cite{hasan-kane-2010} Indeed, we will show that the system takes on
Chern numbers $-3$ through $3$ in the following three subsections. 
\subsection{Gap Closing Momenta and Symmetries}
\begin{figure}
\vspace*{-2em}\hspace*{-2em}\includegraphics{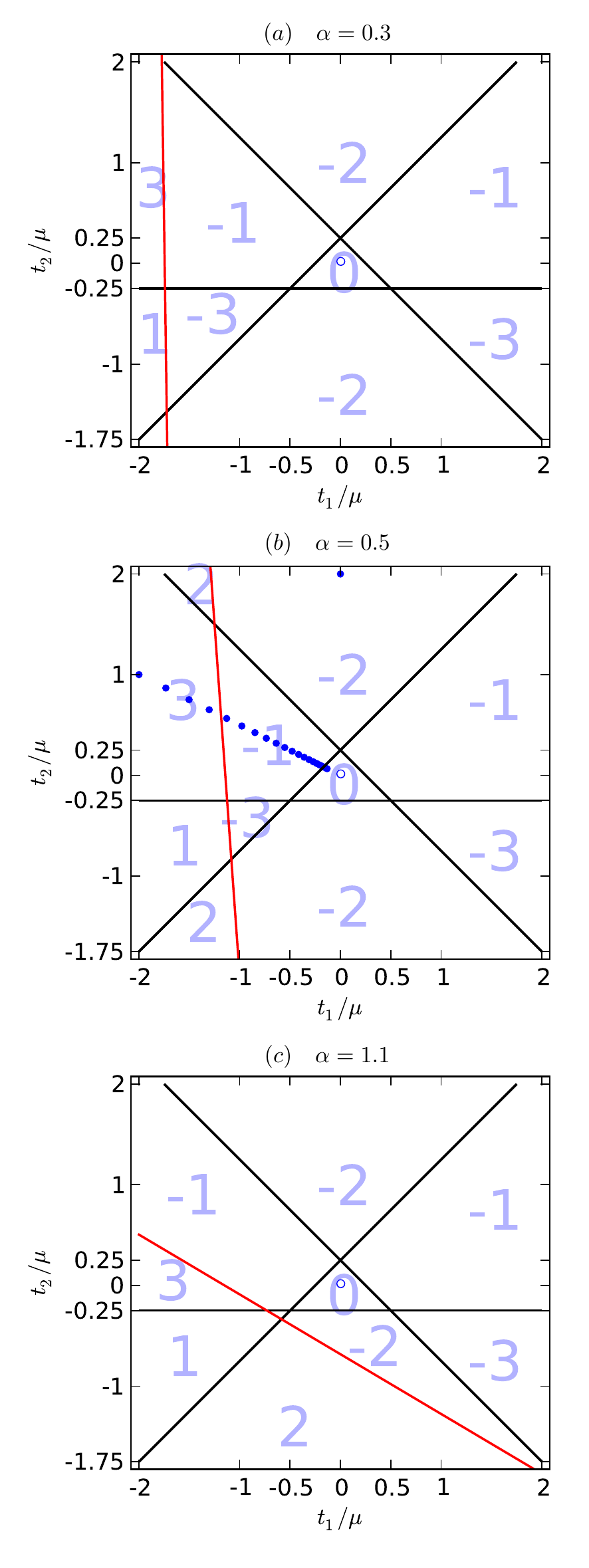}\vspace*{-2em}
\caption{(color online)
Phase diagrams for values of $\alpha=\Delta_1/\Delta_2$, with Chern number for each
phase. The dots subfigure (b) indicate values of $t_1/\mu,t_2/\mu$ investigated numerically.
\label{phase-diagram}}
\end{figure}
In this section, we follow a well-known program for calculating Chern numbers; a brief
review is provided in Appendix \ref{chern-review}. After fourier transforming
the Hamiltonian \ref{hamiltonian}, the bulk band gap is seen to collapse for
momenta $\V k=(k_x,k_y)$ such that
\begin{multline}
0=\frac{1}{2\Delta_2} \Delta(k) = \alpha \lB \sin(k_x)+ i\sin(k_y)\rB \\+\sin(k_x - k_y)+i\sin(k_x + k_y)
\end{multline}
and
\begin{multline}
0=\frac{1}{\mu} h(k) = -1+2\lP \frac{t_1}{\mu}\rP \lB \cos(k_x)+\cos(k_y)\rB+\\4\lP \frac{t_2}{\mu}\rP\cos(k_x)\cos(k_y)  
\end{multline}
Our procedure identifies the zeros of $\Delta(\V k)$ and then characterizes $h(\V k)$
at these momenta.
\subsection{Zeros of \texorpdfstring{$\Delta$}{Delta}}
There are the four zeros of $\Delta$ at the high symmetry points (i.e., where $\sin k_x=\sin k_y=0$).
Assuming that the sine terms do not vanish, the remaining four zeros of $\Delta$ can be shown to satisfy
\begin{align}
\cot(k_x)^2&=\frac{\alpha^2}{2-\alpha^2}\lB 1 + \sqrt{\frac{4}{4+\alpha^4}}\rB \nonumber\\
\cot(k_y)^2&=\frac{\alpha^2}{2-\alpha^2}\lB 1 - \sqrt{\frac{4}{4+\alpha^4}}\rB \label{nndelta_vortices:cot}
\end{align}
Thus, there are two cases: $\alpha^2\geq 2$, in which $\Delta(\V k)$ only vanishes
at the four high-symmetry points, and $\alpha^2<2$, for which the order parameter vanishes
at two additional, $\alpha$-dependent momenta. A straightforward, if lengthy, consideration of cases
of the signs of the $\cos(k_i)$ and $\sin(k_i)$ would allow the cotangent terms to be plotted implicitly, giving 
an exact solution for the location in the Brillouin zone for each zero. Fortunately, an explicit solution for the
momenta of the zeros is not needed, as we will see momentarily.\par
\subsection{\texorpdfstring{$h$}{h} at the zeros of \texorpdfstring{$\Delta$}{Delta}}
For $\V k$ such that  $\Delta(\V k)=0$, the band gap closes if and only if $h(\V k)=0$. 
Furthermore, the phase winding of $\Delta$ around its zeros and the sign of $h(\V k)$ at each zero
indicate the Chern number. This well-known result is reviewed in Appendix \ref{chern-review}.
At the high-symmetry points $\Gamma$, $X$, $Y$, and $M$,
\begin{equation}
\frac{1}{4\mu} \mtrx h(0,0)\\h(\pm \pi,0)\\ h(0,\pm \pi)\\ h(\pm\pi,\pm\pi)\emtrx = \mtrx t_2/\mu+t_1/\mu - \frac{1}{4} \\ -t_2/\mu- \frac{1}{4} \\ -t_2/\mu -\frac{1}{4} \\ t_2/\mu- t_1/\mu - \frac{1}{4} \emtrx 
\end{equation}
By setting $h(\V k)=0$, we get
the three alpha-independent phase transition lines: $t_2/\mu=-1/4$,
$t_2/\mu+ t_1/\mu=1/4$, and $ t_2/\mu-t_1/\mu=1/4$. The Chern number changes
by $1$ when crossing each line (except for the $ t_1/\mu=-1/4$ double line, where the change is $2$).
At the $\alpha$-dependent zeros of $\Delta$ given by Equation (\ref{nndelta_vortices:cot}), $h$
is evaluated (a tedious but straightforward considerations of cases):
\begin{align}
\frac{1}{\mu} h(\V k)
&= -1-\lP \frac{t_1}{\mu} \rP \alpha (2-\alpha^2)-\lP \frac{t_2}{\mu} \rP \alpha^4 
\end{align}
The condition that the $\alpha$-dependent zeros are included or excluded
by the Fermi surface (i.e., $h(\V k)\lessgtr 0$) is recast by defining
\begin{align}
\V \beta = \mtrx -\alpha(2-\alpha^2) \\ -\alpha^4 \emtrx\pad{and} \V t = \frac{1}{\mu} \mtrx t_1\\ t_2\emtrx
\end{align}
and the above condition can be restated as
\begin{align}0\leq \frac{1}{\mu} h(\V k) = -1 +\V t\cdot\V \beta \pad{or} \frac{1}{\beta} \leq \V t\cdot\Vhat \beta \label{unpinned-line}\end{align}
I.e., there is a phase transition line, with closest approach to the $t_1$-$t_2$ origin
given by $Z=\frac{\Vhat \beta}{\beta}$.
Plotting these four phase transition lines, identifying the topologically trivial phase
where where $(t_1,t_2)=(0,0)$, and counting the number of lines crossed allows for the creation
of phase diagrams for various values of $\alpha$, such as those in FIG.~\ref{phase-diagram}.
\section{Defects and Magnetic Fields\label{defects}}
\subsection{Magnetic Fields: Flux Tubes and Vortices}
Here, we explore the response of the superconductor to magnetic fields. We assume we are in an extreme type-II
limit: flux tubes form creating real-space vortices in the superconducting order parameter.
In the two dimensional case at hand, the associated response currents are essentially two-dimensional
and therefore very weak. The natural simplifying limit is to take the London penetration depth
$\lambda\to\infty$ and neglect the response magnetic field. We therefore assume a constant,
unaffected, external magnetic field. Notwithstanding the infinite penetration depth, we still
keep the superconducting coherence length $\xi$ finite, allowing vortices in the superconducting
order parameter. The vortices are therefore localized regions of vanishing superconducting order parameter
$\Delta$, without associated magnetic inhomogeneity, which we now describe more precisely.\par
\begin{figure*}[t]
\hspace*{-3.2em}\includegraphics{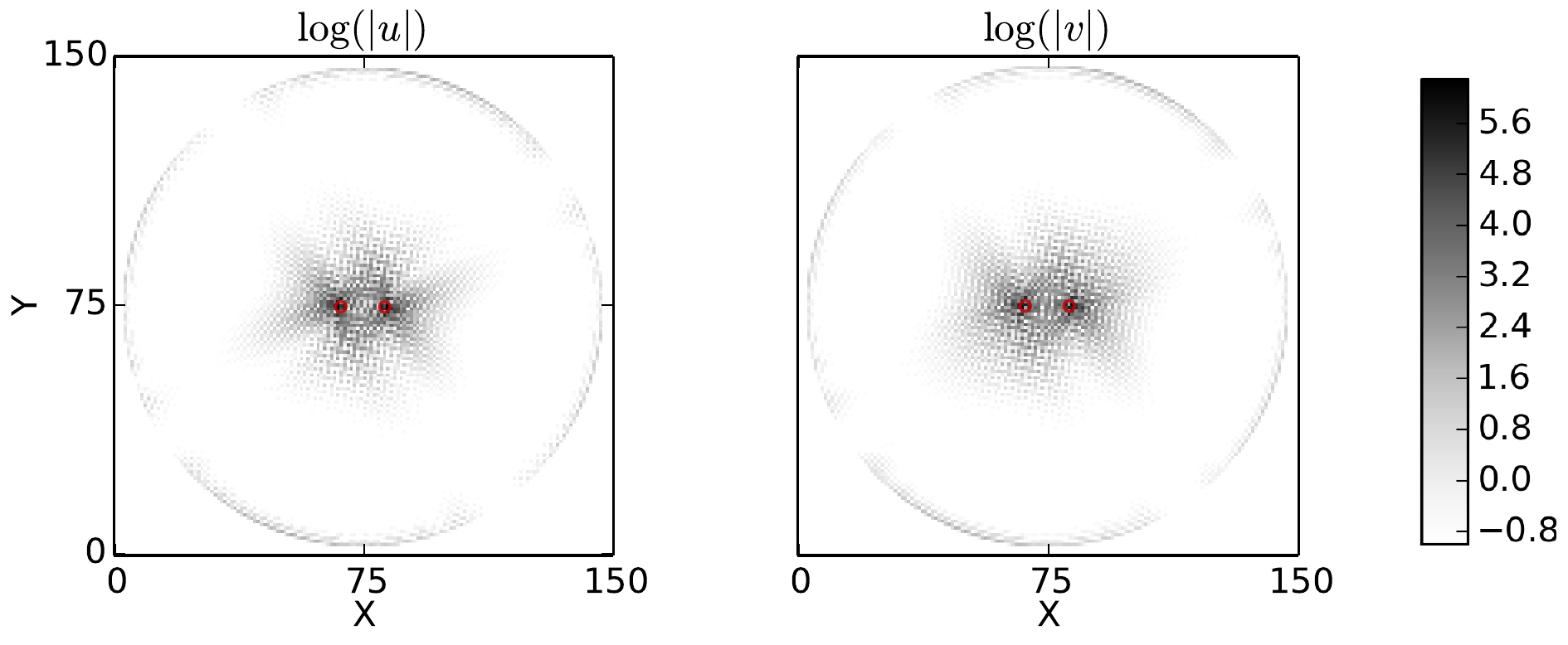}
\caption{(color online) 
Plot of the wavefunction of a vortex mode. The lattice is $150$ by $150$, $t_1=-2$, $\mu=1$,
and $(\Delta_1,\Delta_2)=(0.5,1.0)$ (energy given in units of $t_2$).
The plots are of the natural logarithm of the probability densities of
the $\genfrac{[}{]}{0pt}{}{u}{v}$ parts of the BdG wavefunction. 
The Hamiltonian includes two vortices of radius $r_V=1.6$ indicated by red circles,
separated by $13.2$ in the $x$-direction. The eigenenergy is in-gap: $E/t_2 \approx 5.3\times 10^{-3}$.
\label{vortexmode-illustration}}
\end{figure*}
\begin{figure*}[t]
\hspace*{-3.2em}\includegraphics{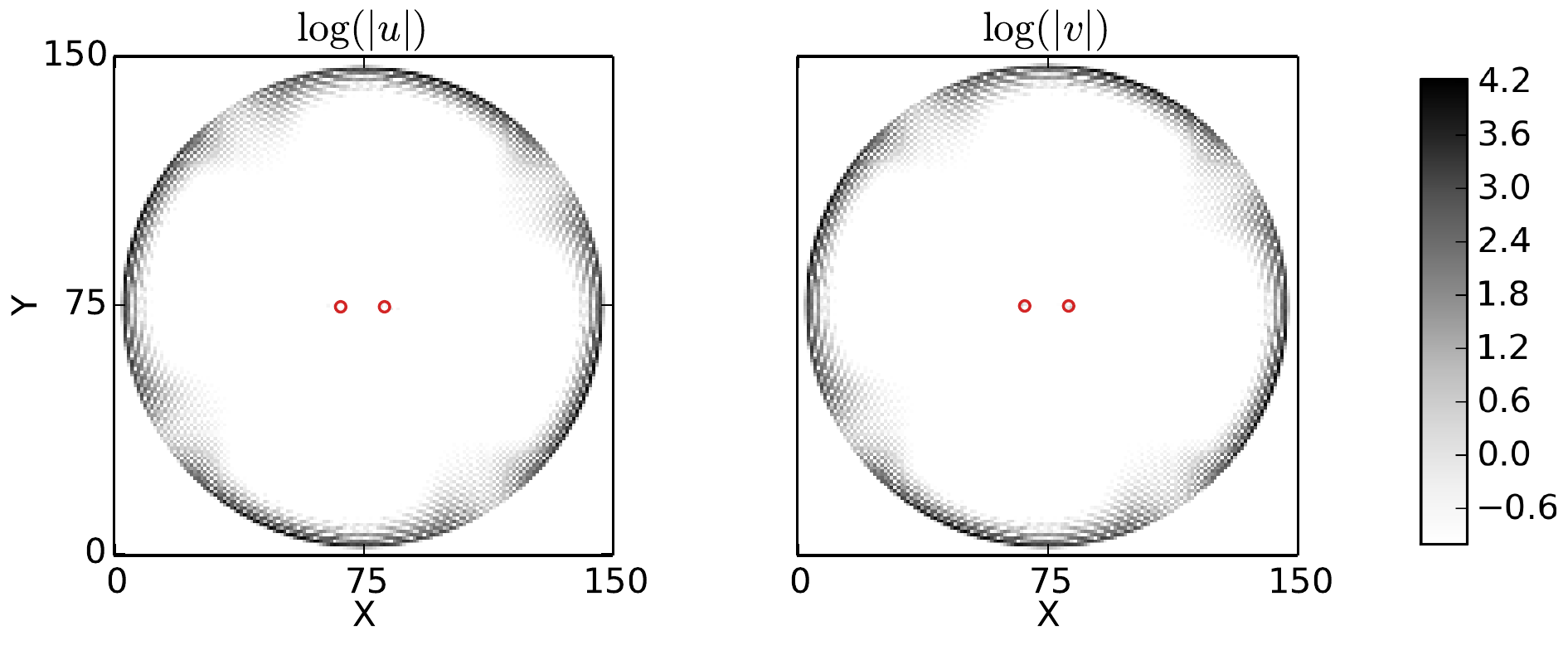}
\caption{(color online) Plot of the wavefunction of an edge mode; compare with FIG.~\ref{vortexmode-illustration}.
The eigenenergy is in-gap: $E/t_2\approx 3.1\times 10^{-3}$.
The state hybridized weakly with the vortices.
\label{edgemode-illustration}}
\end{figure*}
We are guided by the relation
\begin{equation}
 v_s = \frac{1}{2m} \lP\nabla \phi - \frac{2e}{c} \V A\rP \label{semiclassical-phase}
\end{equation}
where $v_s$ is the superfluid velocity, $\phi$ is the phase of the superconducting order
parameter, and $\V A$ is the vector potential, in London gauge. When far away
from a vortex ($r\gg \lambda$), we assume $v_s=0$ and $\V B=0$. Integrating around the vortex yields
\[2\pi n=\oint \nabla \phi \cdot d\V l =  \frac{2e}{c} \Phi_m\]
(i.e., the well-known fact that an integer multiple of magnetic flux quanta penetrates through
a flux tube).  
Because the order parameter is nonzero away from vortices, even for $r < \lambda$, the winding
is an integer multiple of $2\pi$ around each vortex. \par
With a qualitative description of the behavior of the order parameter (the magnitude falls off
near vortices, and the phase winds an integer multiple of $2\pi$ around each vortex), a quantitative
model to perform a numerical simulation must now be established. We use the model\cite{Melikyan_Tesanovic_2006,Vafek_Melikyan_2006}
\begin{equation}
 \Delta_{jk} = \Delta^{(0)}_{k-j} \mathcal{D}\lP j,k\rP e^{i\phi_{jk}}
\end{equation}
The phase of the order parameter $\phi_{jk}$ is a geometric mean
of the expected phases at $j$ and $k$:
\begin{equation}
 e^{i\theta_{jk}} = \frac{e^{i\phi_k}+e^{i\phi_j}}{\lA e^{i\phi_k}+e^{i\phi_j} \rA}
\end{equation}
(The arithmetic mean of $\phi_i$ and $\phi_j$ is insufficient, because the phase for pairing terms crossing any branch cut would
be incorrect.)
Near the vortex cores, $\mathcal{D}$ falls off as
\begin{equation}
 \mathcal{D}(j,k) = \frac{d_\text{eff}(j,k)}{\sqrt{d_\text{eff}(j,k)^2+r_V^2}}
\end{equation}
Where the ``effective distance'' is given by
\begin{equation}
 d_\text{eff}^{-1}(j,k) = \sum_n \lP \min_\text{$x$ between $j$ and $k$} \lA x-v_n\rA\rP^{-1}
\end{equation}
$x$ lies on the line connecting $j$ and $k$.
The vortex core radius $r_V$ is a parameter of the model, on the order of the superconducting coherence length.
Provided that vortices were separated from each other and the edge by many multiples of $r_V$, the vortex core
size was to only weakly affect the measured properties of the system. To reduce the required lattice
sizes for numerical stability, we set $r_V=1.6$, the same order of magnitude as the coherence length in the
cuprates. The hopping terms $h_{jk}$ acquire a Peierls phase due to the magnetic vector potential
\[h_{jk} = h^{(0)}_{k-j}e^{i\frac{e}{c} \int_k^j \V A\cdot d\ell}\]
Relation (\ref{semiclassical-phase}) expresses $\V A$ in London gauge: $\nabla\cdot \V A=0$
and the normal component of $\V A\cdot\hat n$ becomes the physically meaningful
boundary supercurrent. By choosing a gauge where $\V A$ vanishes at the center of
the sample, the vector potential for a constant magnetic field takes the simple form
$\V A\propto \rho\hat\phi$, where $\rho$ is the distance from the center of the sample. Additionally, for
our choice of $\V A$, the boundary current vanishes for circular geometry.
For non-circular geometries, the approximation will remain valid provided that the edge (and associated currents) are
far from the features of interest.
\subsection{Edges}
Square edges can be produced by omitting certain terms in the Hamiltonian, i.e., 
setting all terms of the form $h_{ij}$ and $\Delta_{ij}$ to zero for $ij$ which cross an edge.
While intuitive and simple, when two edges are introduced, an
artificially ``sharp'' corner is produced. The low-energy edge modes that develop
are strongly concentrated at the artificial corners. One might be concerned that
such an unphysical feature might poison the simulation.\par
A choice of smoother edge removes the unphysically sharp corners, but
introduces another problem: there are now lattice sites ``outside'' of the region
of interest. The spectrum will include the unphysical quasiparticle modes outside the edge, 
complicating the analysis. A more natural approach is to make occupation of states beyond
the edge energetically unfavorable. On-site terms $O_i c_i^\dag c_i$
are added with $O_i$ increasingly large near and beyond the edges of the system. For
our purposes, the edge is made very steep and circular, i.e., it goes from $0$ inside
a circular region of the lattice, to a very large number outside it. The lattice sites
with large on-site energies must play no role in the low energy spectrum of the Hamiltonian.
\begin{figure}
\vspace*{-0.5cm}\hspace*{-0.6cm}\includegraphics[trim=0 0 20 0]{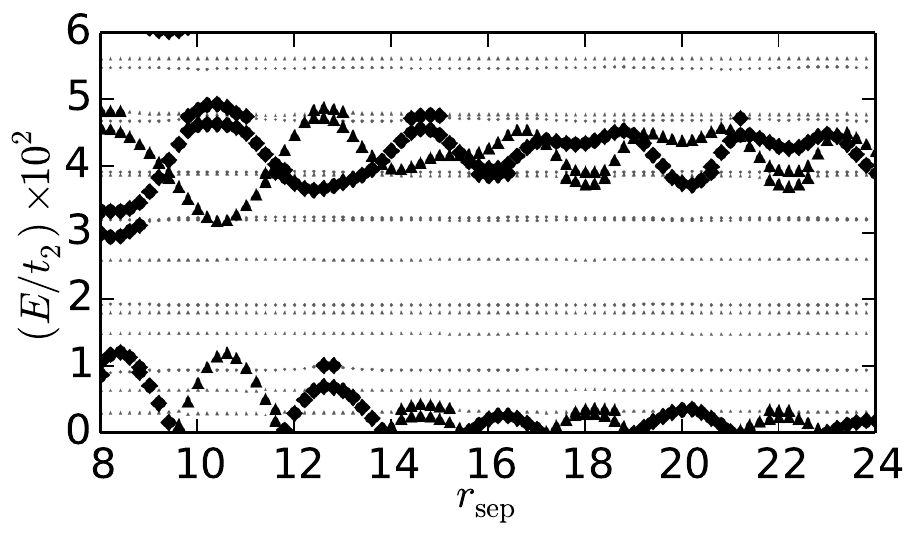}
\caption{Plot of several lowest quasiparticle energies as the separation between two vortices
in the $x$ direction is varied. The lattice was $150$ by $150$, with a circular edge. The other
parameters are $(\Delta_1,\Delta_2)=(0.5,1.0)$, $\mu=1$, and $t_1=-2.0$ (energies are given in terms of $t_2$).
The choice of parameters leads to Chern number $3$ and a single zero-energy vortex core mode, as guaranteed for odd Chern numbers,
c.f. FIG.~\ref{twovortex-separation-chern2}. 
The shapes of the markers indicate the parity of the state under spatial-inversion symmetry: diamond is
even, and triangle is odd (see Appendix \ref{spatial-inversion}). The unimportant edge states are indicated
by the smaller, fainter markers.
\label{twovortex-separation}}
\end{figure}
\subsection{Disorder}
By adjusting the $O_i$ terms, on site disorder is produced, representing quenched impurities on the lattice.
The model is
\[O_i = \lC\begin{array}{llcr} 0,&\text{with probability $1-p$}\\ -E_d,&\text{with probability $p/2$}\\ +E_d,&\text{with probability $p/2$}\end{array}\rN \]
When vortices are moved, such as in FIG.~\ref{disorder-separation}, the same disorder realization
is used for each vortex placement.
\section{Numerical Results\label{results}}
Here, we discuss the results of the numerical diagonalization of the Bogoliubov
de-Gennes Hamiltonian (\ref{hamiltonian}) for eigenvalues near zero. These mid-gap states
arise because of the topological nature of the system. Being deep inside the superconducting
gap, these mid-gap states experience strong particle-hole mixing. As lattice sizes and
vortex separation are increased, hybridization dies off, quasiparticle energies go
to zero, and the particle and hole parts can be made equal, $|u|=|v|$. In our realistic case of finite separation, there will
always be nonzero hybridization, and consequential deviation from equality.\par
We put $(\Delta_1,\Delta_2)=(0.5,1.0)$, and always work with energy is in units of the NNN hopping, $t_2$. Our choice
of parameters creates a rich phase diagram while keeping the magnitudes of both NN and NNN pairing
terms similar. Several choices of $t_1/t_2$ were investigated, but all focus on exploring $t_1<0$
and $t_2>0$, which is similar to the superconducting band of the strontium ruthenates. \par
The output of the numerical simulation is the low energy spectrum and associated wavefunctions. Both
vortex core states, such as FIG.~\ref{vortexmode-illustration}, and edge states, such as 
FIG.~\ref{edgemode-illustration}, are part of the output. Although presented as distinct in the
examples, they can and do hybridize. To distinguish the edge and vortex states automatically,
the probability of a state being present within some distance of the edge is found and used
to classify a given state as ``edge'' or not.
As seen in, for example, FIG.~\ref{twovortex-separation} with $r_\text{sep}\approx 13$,
the edge-vortex hybridization becomes strong enough to cause the third edge modes to hybridize
strongly with the vortex modes, resulting in significant occupation away from the edge.
In general, however, edge modes, being localized away from the vortices (due to the careful choice
of parameters), do not strongly influence the low-energy vortex core modes. 
\begin{figure}
\vspace*{-0.5cm}\hspace*{-0.8cm}\includegraphics{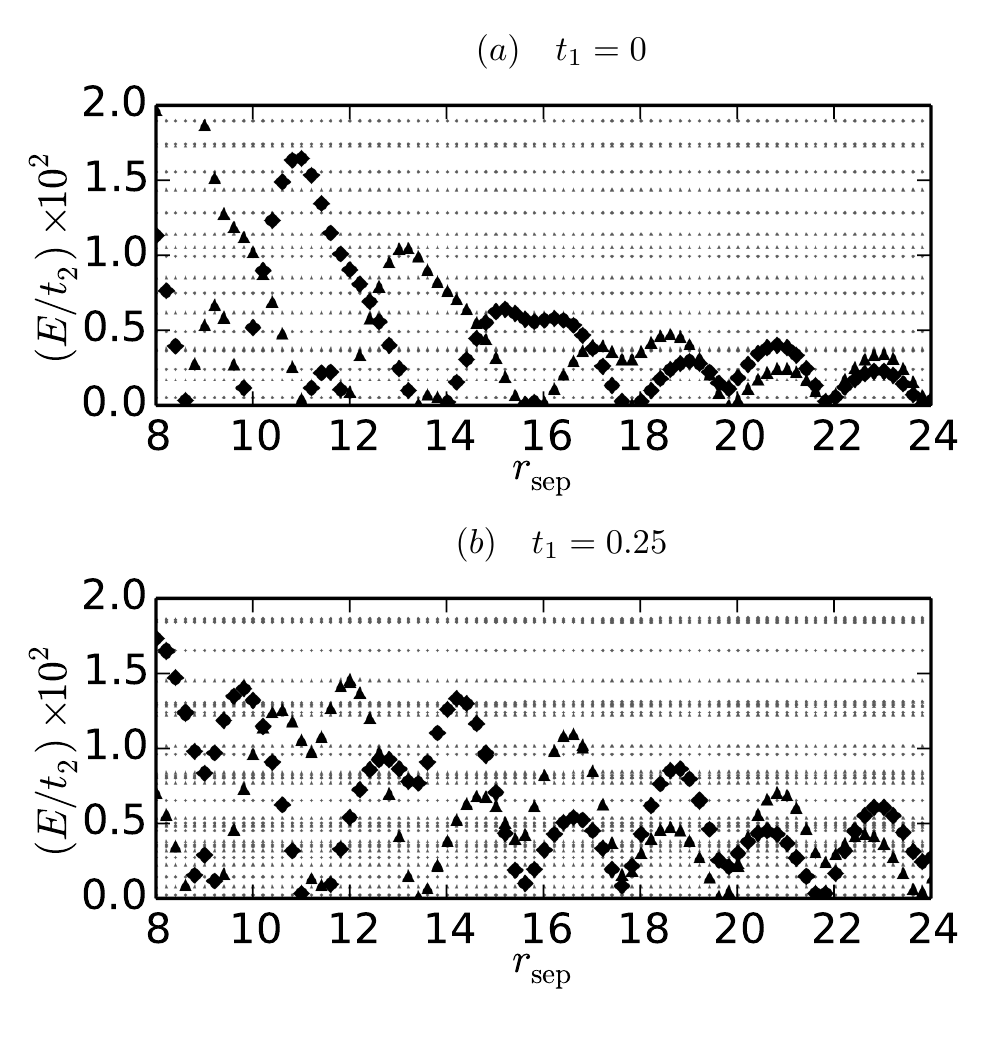}
\caption{Compare with FIG.~\ref{twovortex-separation}: circular lattice, $(\Delta_1,\Delta_2)=(0.25,0.5)$,
and the chemical potential $\mu=0.5$. Each subfigures has a different value of $t_1$, as shown; all energies are given in terms of $t_2$.
The vortex separation $r_\text{sep}$ is given in terms of lattice spacing.
Notice the \emph{two} oscillating low energy excitations, possibly with an
exponentially damped envelope. In the limit of large
separation of the vortices, these vortex core states could become $0$-energy Majorana modes. 
We suspect that that a significant portion of the Chern number $-2$ phase
enjoys these multiple Majorana modes, suggesting analytical investigation. Resistance
to weak disorder is discussed later FIG.~\ref{disorder-separation}.
\label{twovortex-separation-chern2}}
\end{figure}
\begin{figure}
\includegraphics{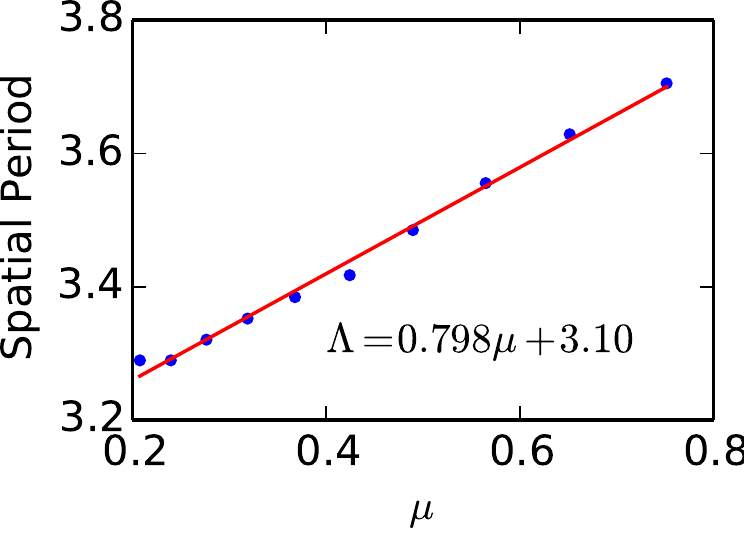}
\caption{(color online) Spatial period of oscillation
of vortex mode energy as two vortices are separated in the $x$ direction, as in FIG.~\ref{twovortex-separation}.
The spatial period changes as the chemical potential $\mu$ is varied (all other parameters are as in the aforementioned
figure). For plotted values of $\mu$, the Chern number was $3$. 
Larger values of $\mu$ were inaccessible due to edge-vortex hybridization.
 \label{mu_spatialperiod}}
\end{figure}
\begin{figure*}[t]
\hspace*{-1.2cm}\includegraphics{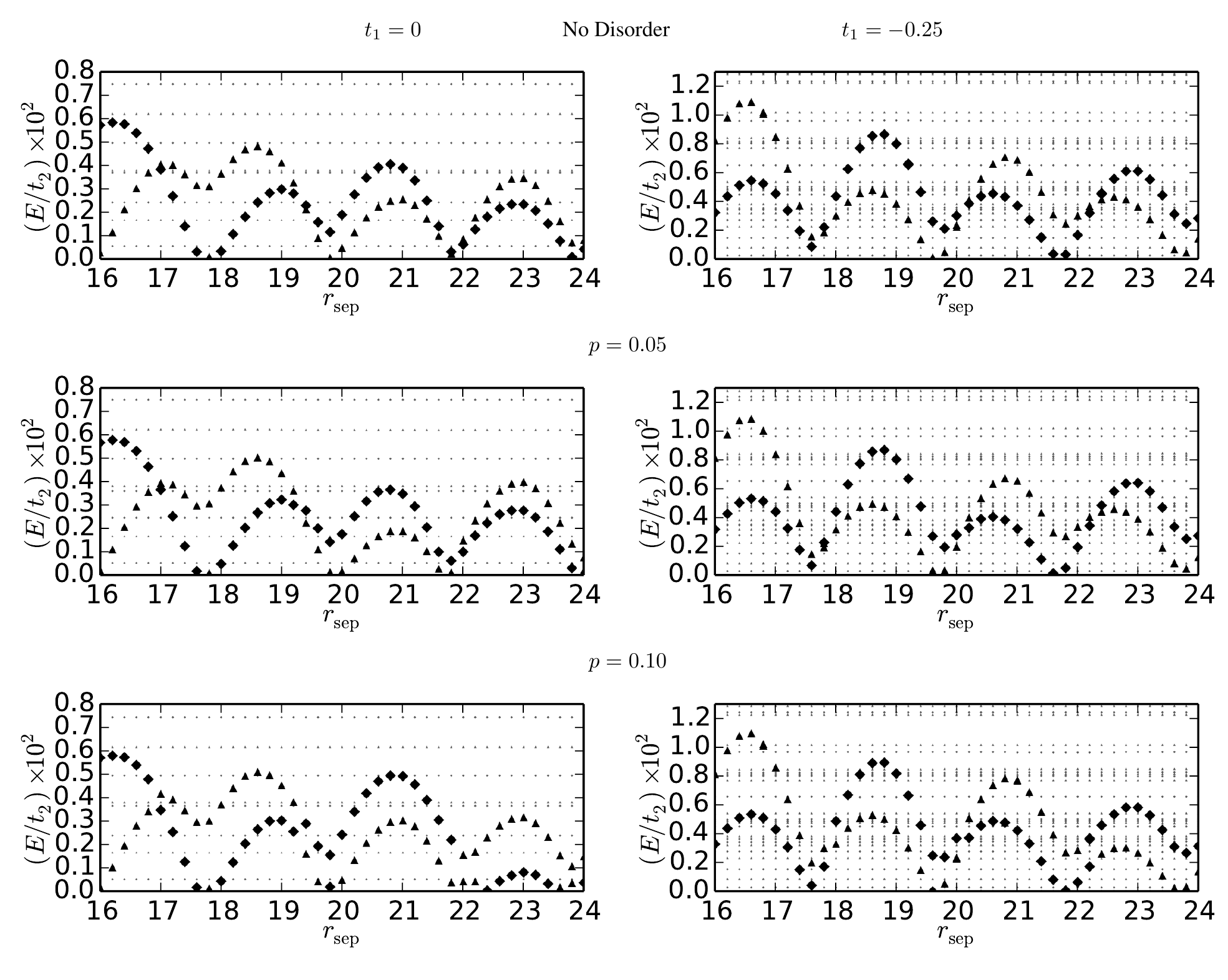}
\caption{Compare with FIG.~\ref{twovortex-separation-chern2}, the choices of parameters is the same same.
The focus of these figures is on the disorder, of strength $E_d = \frac{1}{10}$ and
probability $p$. All energies give in terms of $t_2$.
Although disorder destroys inversion symmetry (discussed in Appendix \ref{spatial-inversion}), there is still significant overlap of wavefunctions with
their spatial inversion partner. The different markers indicate the sign of the overlap: positive is
diamond; negative is triangle; and weak overlap is indicated by a circle. 
Weak disorder \emph{should not} destroy exponentially damped oscillating behavior if it already exists;
no qualitative changes occur when weak disorder is added.
\label{disorder-separation}}
\end{figure*}
\subsection{Vortex Core Mode Oscillations}
When the separation between two magnetic vortices is adjusted, the spectrum shifts,
as seen in FIG.~\ref{twovortex-separation} and FIG.~\ref{twovortex-separation-chern2}.
Most notably, the energies of the lowest quasiparticles exhibit damped oscillation.
The dominant Fourier component of these oscillations is found (and inverted) to give a spatial period. 
The spatial period $\Lambda$ is found as a function of the chemical potential $\mu$ in
FIG.~\ref{mu_spatialperiod} for $t_1=-2$, $\Delta_1=0.5$ (energy given in terms of $t_2$).
The only values of $\mu$ shown are where the vortex core mode only hybridized weakly with the edge modes. 
Even small distortions to the oscillations disturb the calculation of the spatial period significantly.
Systems close to criticality were therefore not examined. In particular, only points in the Chern
number $3$ phase were far enough from criticality to be calculated reliably. In that region,
$\Lambda$ was found to depend linearly on $\mu$, with slope close to $0.8$. As mentioned
before, these oscillations have been analytically\cite{Lutchyn_Galitski_Das_Sarma_2009}
and numerically\cite{mizushima-machida-2010} investigated before (for slightly different systems) with
a period $\sim \frac{2\pi}{k_F}$, due to the oscillations in the vortex mode wavefunctions on the
same spatial period.
\subsection{Majorana Mode Count}
For Chern number $3$, only one vortex mode exists; see FIG.~\ref{twovortex-separation}.
However, for Chern number $\pm 2$, two $0$-energy modes develop when two vortices are introduced.
Modes in the $-2$ region near $t_1=0$ are examined in FIG.~\ref{twovortex-separation-chern2}.
The Chern number $+2$ region (with, e.g., $(t_1/\mu,t_2/\mu)=(-1.5,2)$ in FIG.~\ref{phase-diagram}) 
probably also supports an additional vortex mode, but the issue there is complicated by the fact
that the system is usually quite close to criticality. I.e., the zeros of $\Delta(k)$ occur where $h_k$ is
relatively small, leading to a divergence of correlation lengths. Effective analysis requires
that the edge-vortex hybridization be suppressed; much larger systems would have to be simulated.
\subsection{Disorder} 
Here, we discuss the results on-site disorder to $O_i$ mentioned earlier.
The same simulations with disorder added are shown in FIG.~\ref{disorder-separation}. For weak
disorder, pairs of vortex modes that exist without disorder persist after turning on the weak disorder. In reality, vortices would
become pinned to disorder sites. A more detailed calculation would not install vortices at prespecified locations.
Despite these caveats, we believe that these additional modes warrant further analytical investigation.
\section{Conclusion}
Chiral $p$-wave superconductors on a lattice support additional, interesting phases beyond the two
well-known (topologically trivial) BEC and (Chern number $1$) BCS phases. The Chern number $\pm 2$ phases can be understood
intuitively as a pair of weakly interacting sublattices: the defect modes appear
to survive variation of parameters as well as the addition of weak disorder. It is expected\cite{roy-2010} that some
perturbation of the Hamiltonian will hybridize the defect states, though the precise form of the interaction
has not been determined. The Chern number $3$ phase, on the other hand, does not support any additional modes.
The consequences of including NNN interactions in two-dimensional chiral superconductors are worthy of analytical attention.
\begin{acknowledgments}
 The authors thank Suk Bum Chung, Srinivas Raghu, Rahul Roy, Ipsita Mandal, and Chen-Hsuan Hsu for comments.
S. C. and A. R. were supported by US NSF under the Grant DMR-1004520. Numerical
calculations were performed using Python, Sagemath, and SciPy.
\end{acknowledgments}
\appendix
\section{Chern Number Review\label{chern-review}}
Here, we review Chern number calculation in the defect free case, as in section \ref{analytical} for nearest and next-nearest interactions.
In momentum space, the Hamiltonian (\ref{hamiltonian}) $\mathcal{H}_{ij}$ becomes $\mathcal{H}_k=\V n_k\cdot \tau$, where $\tau$ is a vector of Pauli matrices and
\begin{equation}
 \V n_k = \mtrx \Re \Delta_k \\ -\Im \Delta_k \\ h_k \emtrx.
\end{equation}
he Chern number is obtained by integrating the Berry curvature 
\begin{multline}
\frac{1}{2\pi} \nabla_k \times \bra{0}\Psi_k i\nabla_k \Psi^\dag_k\ket{0}
  = \frac{1}{4\pi } (\nabla_k \Phi_k) \times (\nabla_k n_k)\\
 = \frac{1}{4\pi } \Vhat n \cdot \frac{\partial\Vhat n}{\partial k_x} \times \frac{\partial \Vhat n}{\partial k_y} 
\label{berry-curvature} 
\end{multline}
over the Brillouin zone ($\ket{0}$ is the vacuum state). Both equalities are due to straightforward calculation.
The vector-valued function $\V n$ maps momentum space to $\sR^3$, and characterizes the Cooper pairing (and corresponding
quasiparticles) at a given momentum.
The vanishing of $\V n$ corresponds precisely to nodes in the band structure. Therefore,
in the fully-gapped regime, the unit vector $\Vhat n$ maps $T^2$ to $S^2$, and the above integral
is just the degree of the map $\Vhat n$, an integer.\cite{nakahara-2003} According to the Hopf classification,
the degree characterizes the mapping $\Vhat n$ topologically, i.e. up to homotopy. We emphasize here
that we have so far said nothing about the presence of zero-energy modes or sublattices: only the topologically invariant Chern number.\par
One can do slightly better. By smoothly deforming $\V n$ so that $\V n$ is $\pm\Vhat z$ except when $h_k$ vanishes,
the Chern number is seen to depend only on the winding of the phase of the superconducting order parameter around the Fermi surface (of the parent state, i.e., where $h_k=0$).
Because such a smooth deformation will not close the band gap, the topological invariant is unchanged.
The integral over the Brillouin zone therefore becomes a line integral over the $h_k=0$ surface, which is sensitive only to 
the winding of the superconducting order parameter's phase $\phi$.\par
The winding of $\phi$ can only occurs around zeros of $\Delta$, and always in multiples of $2\pi$.
Neglecting higher-order zeros of $\Delta$, one simply counts the number of zeros enclosed by the Fermi surface, and note whether their winding is clockwise
or counterclockwise to get the Chern number. To get the sign of the answer correct, ``enclosed'' is taken to mean the particle-like side of the Fermi surface.
We emphasize now that we are dealing with a quadratic, \emph{single-band} Hamiltonian. Analogous results
for multi-band Hamiltonians would be more complicated.
\section{Spatial Inversion Symmetry\label{spatial-inversion}}
In the absence of added disorder,  the model systems we consider have a spatial
inversion symmetry which can be exploited to enhance the clarity of some plots
(especially in highlighting oscillatory behavior).
The hopping terms straightforwardly satisfy $h_{-i,-j}=h_{ij}$, while
the $p$-wave symmetry of the superconducting order parameter implies $\Delta_{-i,-j}=-\Delta_{ij}$. Let $\mathcal{I}$ realize the inversion symmetry in
position space (i.e., $(\mathcal{I} h)_{ij}=h_{-i,-j}$) and put $\op I = \tau_3\otimes\mathcal{I}$ ($\tau_3$
acts on the space of Nambu spinors). Clearly, $\op I^2=1$, and $\op I^\dag=\op I$, so the
eigenvalues of $\op I$ are $\pm 1$. The inversion symmetry's action on the Hamiltonian,
\begin{multline}
\op I^\dag \mathcal{H}_{ij} \op I = \mathcal{I}^\dag  \tau_3^\dag \mathcal{H}_{ij} \tau_3\mathcal{I} \\
\mathcal{I}^\dag \mtrx h_{ij} & -\Delta_{ij} \\ -\bar \Delta_{ji} & -\bar h_{ij} \emtrx \mathcal{I}
=  \mtrx h_{-i-j} & -\Delta_{-i-j} \\ -\bar \Delta_{-j-i} & -\bar h_{-i-j} \emtrx \\
= \mtrx h_{ij} & \Delta_{ij} \\ \bar \Delta_{ji} & -h_{ij} \emtrx
= \mathcal{H}_{ij}
\end{multline}
shows that the inversion symmetry $\op I$ relates quasiparticles of the form
$(u_i,v_i)$ to $(u_{-i},-v_{-i})$. For appropriate energy eigenstates, $\psi=\pm \op I\psi$ and the subparts $u$ and $v$ therefore
have separate (and opposite) inversion symmetries given by $\mathcal{I} u=\pm u$ and $\mathcal{I} v=\mp v$.
These eigenvalues can be changed using the Bogoliubov-de Gennes particle-hole symmetry (i.e., $\Xi=\tau_1 \otimes K$, where $K$
is complex conjugation); the symmetry-related negative-energy pair has opposite $\op I$ eigenvalue: 
\begin{align}
\Xi \op I &= \tau_1 K \tau_3 \op I=-i\tau_2 \op I=-\op I \Xi 
\end{align}
\par In the case of disorder, the symmetry $\op I$ is clearly broken by the additional terms.
Nonetheless, the overlap $\braket{\psi}{\op I\psi}$ is still meaningful:
if positive, we can still identify $\psi$ as ``symmetric''-like or otherwise.
In the figures, e.g., FIG.~\ref{disorder-separation}, the sign of the overlap is plotted as the shape of the symbol.

\begin{thebibliography}{26}%
\makeatletter
\providecommand \@ifxundefined [1]{%
 \@ifx{#1\undefined}
}%
\providecommand \@ifnum [1]{%
 \ifnum #1\expandafter \@firstoftwo
 \else \expandafter \@secondoftwo
 \fi
}%
\providecommand \@ifx [1]{%
 \ifx #1\expandafter \@firstoftwo
 \else \expandafter \@secondoftwo
 \fi
}%
\providecommand \natexlab [1]{#1}%
\providecommand \enquote  [1]{``#1''}%
\providecommand \bibnamefont  [1]{#1}%
\providecommand \bibfnamefont [1]{#1}%
\providecommand \citenamefont [1]{#1}%
\providecommand \href@noop [0]{\@secondoftwo}%
\providecommand \href [0]{\begingroup \@sanitize@url \@href}%
\providecommand \@href[1]{\@@startlink{#1}\@@href}%
\providecommand \@@href[1]{\endgroup#1\@@endlink}%
\providecommand \@sanitize@url [0]{\catcode `\\12\catcode `\$12\catcode
  `\&12\catcode `\#12\catcode `\^12\catcode `\_12\catcode `\%12\relax}%
\providecommand \@@startlink[1]{}%
\providecommand \@@endlink[0]{}%
\providecommand \url  [0]{\begingroup\@sanitize@url \@url }%
\providecommand \@url [1]{\endgroup\@href {#1}{\urlprefix }}%
\providecommand \urlprefix  [0]{URL }%
\providecommand \Eprint [0]{\href }%
\providecommand \doibase [0]{http://dx.doi.org/}%
\providecommand \selectlanguage [0]{\@gobble}%
\providecommand \bibinfo  [0]{\@secondoftwo}%
\providecommand \bibfield  [0]{\@secondoftwo}%
\providecommand \translation [1]{[#1]}%
\providecommand \BibitemOpen [0]{}%
\providecommand \bibitemStop [0]{}%
\providecommand \bibitemNoStop [0]{.\EOS\space}%
\providecommand \EOS [0]{\spacefactor3000\relax}%
\providecommand \BibitemShut  [1]{\csname bibitem#1\endcsname}%
\let\auto@bib@innerbib\@empty
\bibitem [{\citenamefont {Mourik}\ \emph {et~al.}(2012)\citenamefont {Mourik},
  \citenamefont {Zuo}, \citenamefont {Frolov}, \citenamefont {Plissard},
  \citenamefont {Bakkers},\ and\ \citenamefont
  {Kouwenhoven}}]{kouwenhoven-2012}%
  \BibitemOpen
  \bibfield  {author} {\bibinfo {author} {\bibfnamefont {V.}~\bibnamefont
  {Mourik}}, \bibinfo {author} {\bibfnamefont {K.}~\bibnamefont {Zuo}},
  \bibinfo {author} {\bibfnamefont {S.~M.}\ \bibnamefont {Frolov}}, \bibinfo
  {author} {\bibfnamefont {S.~R.}\ \bibnamefont {Plissard}}, \bibinfo {author}
  {\bibfnamefont {E.~P. A.~M.}\ \bibnamefont {Bakkers}}, \ and\ \bibinfo
  {author} {\bibfnamefont {L.~P.}\ \bibnamefont {Kouwenhoven}},\ }\href
  {\doibase 10.1126/science.1222360} {\ \textbf {\bibinfo {volume} {336}},\
  \bibinfo {pages} {1003} (\bibinfo {year} {2012})}\BibitemShut {NoStop}%
\bibitem [{\citenamefont {Kitaev}(2001)}]{kitaev-2001}%
  \BibitemOpen
  \bibfield  {author} {\bibinfo {author} {\bibfnamefont {A.~Y.}\ \bibnamefont
  {Kitaev}},\ }\href {http://stacks.iop.org/1063-7869/44/i=10S/a=S29}
  {\bibfield  {journal} {\bibinfo  {journal} {Physics-Uspekhi}\ }\textbf
  {\bibinfo {volume} {44}},\ \bibinfo {pages} {131} (\bibinfo {year}
  {2001})}\BibitemShut {NoStop}%
\bibitem [{\citenamefont {Wilczek}(2009)}]{wilczek-2009}%
  \BibitemOpen
  \bibfield  {author} {\bibinfo {author} {\bibfnamefont {F.}~\bibnamefont
  {Wilczek}},\ }\href {\doibase 10.1038/nphys1380} {\bibfield  {journal}
  {\bibinfo  {journal} {Nat Phys}\ }\textbf {\bibinfo {volume} {5}},\ \bibinfo
  {pages} {614} (\bibinfo {year} {2009})}\BibitemShut {NoStop}%
\bibitem [{\citenamefont {Moore}\ and\ \citenamefont
  {Read}(1991)}]{moore-read-1991}%
  \BibitemOpen
  \bibfield  {author} {\bibinfo {author} {\bibfnamefont {G.}~\bibnamefont
  {Moore}}\ and\ \bibinfo {author} {\bibfnamefont {N.}~\bibnamefont {Read}},\
  }\href {\doibase 10.1016/0550-3213(91)90407-O} {\bibfield  {journal}
  {\bibinfo  {journal} {Nuclear Physics B}\ }\textbf {\bibinfo {volume}
  {360}},\ \bibinfo {pages} {362} (\bibinfo {year} {1991})}\BibitemShut
  {NoStop}%
\bibitem [{\citenamefont {Fu}\ and\ \citenamefont {Kane}(2008)}]{fu-kane-2008}%
  \BibitemOpen
  \bibfield  {author} {\bibinfo {author} {\bibfnamefont {L.}~\bibnamefont
  {Fu}}\ and\ \bibinfo {author} {\bibfnamefont {C.~L.}\ \bibnamefont {Kane}},\
  }\href {\doibase 10.1103/PhysRevLett.100.096407} {\bibfield  {journal}
  {\bibinfo  {journal} {Phys. Rev. Lett.}\ }\textbf {\bibinfo {volume} {100}},\
  \bibinfo {pages} {096407} (\bibinfo {year} {2008})}\BibitemShut {NoStop}%
\bibitem [{\citenamefont {Sau}\ \emph {et~al.}(2010{\natexlab{a}})\citenamefont
  {Sau}, \citenamefont {Lutchyn}, \citenamefont {Tewari},\ and\ \citenamefont
  {Das~Sarma}}]{sau-lutchyn-2010}%
  \BibitemOpen
  \bibfield  {author} {\bibinfo {author} {\bibfnamefont {J.~D.}\ \bibnamefont
  {Sau}}, \bibinfo {author} {\bibfnamefont {R.~M.}\ \bibnamefont {Lutchyn}},
  \bibinfo {author} {\bibfnamefont {S.}~\bibnamefont {Tewari}}, \ and\ \bibinfo
  {author} {\bibfnamefont {S.}~\bibnamefont {Das~Sarma}},\ }\href {\doibase
  10.1103/PhysRevLett.104.040502} {\bibfield  {journal} {\bibinfo  {journal}
  {Phys. Rev. Lett.}\ }\textbf {\bibinfo {volume} {104}},\ \bibinfo {pages}
  {040502} (\bibinfo {year} {2010}{\natexlab{a}})}\BibitemShut {NoStop}%
\bibitem [{\citenamefont {Sau}\ \emph {et~al.}(2010{\natexlab{b}})\citenamefont
  {Sau}, \citenamefont {Tewari}, \citenamefont {Lutchyn}, \citenamefont
  {Stanescu},\ and\ \citenamefont {Das~Sarma}}]{sau-tewari-2010}%
  \BibitemOpen
  \bibfield  {author} {\bibinfo {author} {\bibfnamefont {J.~D.}\ \bibnamefont
  {Sau}}, \bibinfo {author} {\bibfnamefont {S.}~\bibnamefont {Tewari}},
  \bibinfo {author} {\bibfnamefont {R.~M.}\ \bibnamefont {Lutchyn}}, \bibinfo
  {author} {\bibfnamefont {T.~D.}\ \bibnamefont {Stanescu}}, \ and\ \bibinfo
  {author} {\bibfnamefont {S.}~\bibnamefont {Das~Sarma}},\ }\href {\doibase
  10.1103/PhysRevB.82.214509} {\bibfield  {journal} {\bibinfo  {journal} {Phys.
  Rev. B}\ }\textbf {\bibinfo {volume} {82}},\ \bibinfo {pages} {214509}
  (\bibinfo {year} {2010}{\natexlab{b}})}\BibitemShut {NoStop}%
\bibitem [{\citenamefont {Alicea}\ \emph {et~al.}(2011)\citenamefont {Alicea},
  \citenamefont {Oreg}, \citenamefont {Refael}, \citenamefont {von Oppen},\
  and\ \citenamefont {Fisher}}]{g_Refael_von_Oppen_Fisher_2011}%
  \BibitemOpen
  \bibfield  {author} {\bibinfo {author} {\bibfnamefont {J.}~\bibnamefont
  {Alicea}}, \bibinfo {author} {\bibfnamefont {Y.}~\bibnamefont {Oreg}},
  \bibinfo {author} {\bibfnamefont {G.}~\bibnamefont {Refael}}, \bibinfo
  {author} {\bibfnamefont {F.}~\bibnamefont {von Oppen}}, \ and\ \bibinfo
  {author} {\bibfnamefont {M.~P.~A.}\ \bibnamefont {Fisher}},\ }\href {\doibase
  10.1038/nphys1915} {\bibfield  {journal} {\bibinfo  {journal} {Nature
  Physics}\ }\textbf {\bibinfo {volume} {7}},\ \bibinfo {pages} {412} (\bibinfo
  {year} {2011})}\BibitemShut {NoStop}%
\bibitem [{\citenamefont {Alicea}(2010)}]{alicea-2010}%
  \BibitemOpen
  \bibfield  {author} {\bibinfo {author} {\bibfnamefont {J.}~\bibnamefont
  {Alicea}},\ }\href {\doibase 10.1103/PhysRevB.81.125318} {\bibfield
  {journal} {\bibinfo  {journal} {Phys. Rev. B}\ }\textbf {\bibinfo {volume}
  {81}},\ \bibinfo {pages} {125318} (\bibinfo {year} {2010})}\BibitemShut
  {NoStop}%
\bibitem [{\citenamefont {Lutchyn}, \citenamefont {Sau},\ and\ \citenamefont
  {Das~Sarma}(2010)}]{lutchyn-sau-2010}%
  \BibitemOpen
  \bibfield  {author} {\bibinfo {author} {\bibfnamefont {R.~M.}\ \bibnamefont
  {Lutchyn}}, \bibinfo {author} {\bibfnamefont {J.~D.}\ \bibnamefont {Sau}}, \
  and\ \bibinfo {author} {\bibfnamefont {S.}~\bibnamefont {Das~Sarma}},\ }\href
  {\doibase 10.1103/PhysRevLett.105.077001} {\bibfield  {journal} {\bibinfo
  {journal} {Phys. Rev. Lett.}\ }\textbf {\bibinfo {volume} {105}},\ \bibinfo
  {pages} {077001} (\bibinfo {year} {2010})}\BibitemShut {NoStop}%
\bibitem [{\citenamefont {DeGottardi}, \citenamefont {Sen},\ and\ \citenamefont
  {Vishveshwara}(2011)}]{degottardi-diptiman-2011}%
  \BibitemOpen
  \bibfield  {author} {\bibinfo {author} {\bibfnamefont {W.}~\bibnamefont
  {DeGottardi}}, \bibinfo {author} {\bibfnamefont {D.}~\bibnamefont {Sen}}, \
  and\ \bibinfo {author} {\bibfnamefont {S.}~\bibnamefont {Vishveshwara}},\
  }\href {\doibase 10.1088/1367-2630/13/6/065028} {\bibfield  {journal}
  {\bibinfo  {journal} {New Journal of Physics}\ }\textbf {\bibinfo {volume}
  {13}},\ \bibinfo {pages} {065028} (\bibinfo {year} {2011})}\BibitemShut
  {NoStop}%
\bibitem [{\citenamefont {Alicea}(2012)}]{Alicea_2012}%
  \BibitemOpen
  \bibfield  {author} {\bibinfo {author} {\bibfnamefont {J.}~\bibnamefont
  {Alicea}},\ }\href {\doibase 10.1088/0034-4885/75/7/076501} {\bibfield
  {journal} {\bibinfo  {journal} {Reports on Progress in Physics}\ }\textbf
  {\bibinfo {volume} {75}},\ \bibinfo {pages} {076501} (\bibinfo {year}
  {2012})}\BibitemShut {NoStop}%
\bibitem [{\citenamefont {Read}\ and\ \citenamefont
  {Green}(2000)}]{read-green-2000}%
  \BibitemOpen
  \bibfield  {author} {\bibinfo {author} {\bibfnamefont {N.}~\bibnamefont
  {Read}}\ and\ \bibinfo {author} {\bibfnamefont {D.}~\bibnamefont {Green}},\
  }\href {\doibase 10.1103/PhysRevB.61.10267} {\bibfield  {journal} {\bibinfo
  {journal} {Phys. Rev. B}\ }\textbf {\bibinfo {volume} {61}},\ \bibinfo
  {pages} {10267} (\bibinfo {year} {2000})}\BibitemShut {NoStop}%
\bibitem [{\citenamefont {Ivanov}(2001)}]{ivanov-2001}%
  \BibitemOpen
  \bibfield  {author} {\bibinfo {author} {\bibfnamefont {D.~A.}\ \bibnamefont
  {Ivanov}},\ }\href {\doibase 10.1103/PhysRevLett.86.268} {\bibfield
  {journal} {\bibinfo  {journal} {Phys. Rev. Lett.}\ }\textbf {\bibinfo
  {volume} {86}},\ \bibinfo {pages} {268} (\bibinfo {year} {2001})}\BibitemShut
  {NoStop}%
\bibitem [{\citenamefont {Roy}(2010)}]{roy-2010}%
  \BibitemOpen
  \bibfield  {author} {\bibinfo {author} {\bibfnamefont {R.}~\bibnamefont
  {Roy}},\ }\href {\doibase 10.1103/PhysRevLett.105.186401} {\bibfield
  {journal} {\bibinfo  {journal} {Phys. Rev. Lett.}\ }\textbf {\bibinfo
  {volume} {105}},\ \bibinfo {pages} {186401} (\bibinfo {year}
  {2010})}\BibitemShut {NoStop}%
\bibitem [{\citenamefont {Kopnin}\ and\ \citenamefont
  {Salomaa}(1991)}]{kopnin-salomaa-1991}%
  \BibitemOpen
  \bibfield  {author} {\bibinfo {author} {\bibfnamefont {N.~B.}\ \bibnamefont
  {Kopnin}}\ and\ \bibinfo {author} {\bibfnamefont {M.~M.}\ \bibnamefont
  {Salomaa}},\ }\href {\doibase 10.1103/PhysRevB.44.9667} {\bibfield  {journal}
  {\bibinfo  {journal} {Phys. Rev. B}\ }\textbf {\bibinfo {volume} {44}},\
  \bibinfo {pages} {9667} (\bibinfo {year} {1991})}\BibitemShut {NoStop}%
\bibitem [{\citenamefont {Volovik}(1999)}]{volovik-1999}%
  \BibitemOpen
  \bibfield  {author} {\bibinfo {author} {\bibfnamefont {G.}~\bibnamefont
  {Volovik}},\ }\href {\doibase 10.1134/1.568223} {\bibfield  {journal}
  {\bibinfo  {journal} {Journal of Experimental and Theoretical Physics
  Letters}\ }\textbf {\bibinfo {volume} {70}},\ \bibinfo {pages} {609}
  (\bibinfo {year} {1999})}\BibitemShut {NoStop}%
\bibitem [{\citenamefont {Caroli}, \citenamefont {Gennes},\ and\ \citenamefont
  {Matricon}(1964)}]{caroli-1964}%
  \BibitemOpen
  \bibfield  {author} {\bibinfo {author} {\bibfnamefont {C.}~\bibnamefont
  {Caroli}}, \bibinfo {author} {\bibfnamefont {P.~D.}\ \bibnamefont {Gennes}},
  \ and\ \bibinfo {author} {\bibfnamefont {J.}~\bibnamefont {Matricon}},\
  }\href {\doibase 10.1016/0031-9163(64)90375-0} {\bibfield  {journal}
  {\bibinfo  {journal} {Physics Letters}\ }\textbf {\bibinfo {volume} {9}},\
  \bibinfo {pages} {307 } (\bibinfo {year} {1964})}\BibitemShut {NoStop}%
\bibitem [{\citenamefont {Hasan}\ and\ \citenamefont
  {Kane}(2010)}]{hasan-kane-2010}%
  \BibitemOpen
  \bibfield  {author} {\bibinfo {author} {\bibfnamefont {M.~Z.}\ \bibnamefont
  {Hasan}}\ and\ \bibinfo {author} {\bibfnamefont {C.~L.}\ \bibnamefont
  {Kane}},\ }\href {\doibase 10.1103/RevModPhys.82.3045} {\bibfield  {journal}
  {\bibinfo  {journal} {Rev. Mod. Phys.}\ }\textbf {\bibinfo {volume} {82}},\
  \bibinfo {pages} {3045} (\bibinfo {year} {2010})}\BibitemShut {NoStop}%
\bibitem [{\citenamefont {Niu}\ \emph {et~al.}(2012)\citenamefont {Niu},
  \citenamefont {Chung}, \citenamefont {Hsu}, \citenamefont {Mandal},
  \citenamefont {Raghu},\ and\ \citenamefont
  {Chakravarty}}]{niu-chakravarty-2012}%
  \BibitemOpen
  \bibfield  {author} {\bibinfo {author} {\bibfnamefont {Y.}~\bibnamefont
  {Niu}}, \bibinfo {author} {\bibfnamefont {S.~B.}\ \bibnamefont {Chung}},
  \bibinfo {author} {\bibfnamefont {C.-H.}\ \bibnamefont {Hsu}}, \bibinfo
  {author} {\bibfnamefont {I.}~\bibnamefont {Mandal}}, \bibinfo {author}
  {\bibfnamefont {S.}~\bibnamefont {Raghu}}, \ and\ \bibinfo {author}
  {\bibfnamefont {S.}~\bibnamefont {Chakravarty}},\ }\href {\doibase
  10.1103/PhysRevB.85.035110} {\bibfield  {journal} {\bibinfo  {journal} {Phys.
  Rev. B}\ }\textbf {\bibinfo {volume} {85}},\ \bibinfo {pages} {035110}
  (\bibinfo {year} {2012})}\BibitemShut {NoStop}%
\bibitem [{\citenamefont {Cheng}\ \emph {et~al.}(2009)\citenamefont {Cheng},
  \citenamefont {Lutchyn}, \citenamefont {Galitski},\ and\ \citenamefont
  {Das~Sarma}}]{Lutchyn_Galitski_Das_Sarma_2009}%
  \BibitemOpen
  \bibfield  {author} {\bibinfo {author} {\bibfnamefont {M.}~\bibnamefont
  {Cheng}}, \bibinfo {author} {\bibfnamefont {R.}~\bibnamefont {Lutchyn}},
  \bibinfo {author} {\bibfnamefont {V.}~\bibnamefont {Galitski}}, \ and\
  \bibinfo {author} {\bibfnamefont {S.}~\bibnamefont {Das~Sarma}},\ }\href
  {\doibase 10.1103/PhysRevLett.103.107001} {\bibfield  {journal} {\bibinfo
  {journal} {Physical Review Letters}\ }\textbf {\bibinfo {volume} {103}},\
  \bibinfo {pages} {107001} (\bibinfo {year} {2009})}\BibitemShut {NoStop}%
\bibitem [{\citenamefont {Mizushima}\ and\ \citenamefont
  {Machida}(2010)}]{mizushima-machida-2010}%
  \BibitemOpen
  \bibfield  {author} {\bibinfo {author} {\bibfnamefont {T.}~\bibnamefont
  {Mizushima}}\ and\ \bibinfo {author} {\bibfnamefont {K.}~\bibnamefont
  {Machida}},\ }\href {\doibase 10.1103/PhysRevA.82.023624} {\bibfield
  {journal} {\bibinfo  {journal} {Phys. Rev. A}\ }\textbf {\bibinfo {volume}
  {82}},\ \bibinfo {pages} {023624} (\bibinfo {year} {2010})}\BibitemShut
  {NoStop}%
\bibitem [{\citenamefont {Altland}\ and\ \citenamefont
  {Zirnbauer}(1997)}]{altland-zirnbauer-1997}%
  \BibitemOpen
  \bibfield  {author} {\bibinfo {author} {\bibfnamefont {A.}~\bibnamefont
  {Altland}}\ and\ \bibinfo {author} {\bibfnamefont {M.~R.}\ \bibnamefont
  {Zirnbauer}},\ }\href {\doibase 10.1103/PhysRevB.55.1142} {\bibfield
  {journal} {\bibinfo  {journal} {Phys. Rev. B}\ }\textbf {\bibinfo {volume}
  {55}},\ \bibinfo {pages} {1142} (\bibinfo {year} {1997})}\BibitemShut
  {NoStop}%
\bibitem [{\citenamefont {Melikyan}\ and\ \citenamefont
  {Te\v{s}anovi\'{c}}(2006)}]{Melikyan_Tesanovic_2006}%
  \BibitemOpen
  \bibfield  {author} {\bibinfo {author} {\bibfnamefont {A.}~\bibnamefont
  {Melikyan}}\ and\ \bibinfo {author} {\bibfnamefont {Z.}~\bibnamefont
  {Te\v{s}anovi\'{c}}},\ }\href {\doibase 10.1103/PhysRevB.74.144501}
  {\bibfield  {journal} {\bibinfo  {journal} {Phys. Rev. B}\ }\textbf {\bibinfo
  {volume} {74}},\ \bibinfo {pages} {144501} (\bibinfo {year}
  {2006})}\BibitemShut {NoStop}%
\bibitem [{\citenamefont {Vafek}\ and\ \citenamefont
  {Melikyan}(2006)}]{Vafek_Melikyan_2006}%
  \BibitemOpen
  \bibfield  {author} {\bibinfo {author} {\bibfnamefont {O.}~\bibnamefont
  {Vafek}}\ and\ \bibinfo {author} {\bibfnamefont {A.}~\bibnamefont
  {Melikyan}},\ }\href {\doibase 10.1103/PhysRevLett.96.167005} {\bibfield
  {journal} {\bibinfo  {journal} {Physical Review Letters}\ }\textbf {\bibinfo
  {volume} {96}},\ \bibinfo {pages} {167005} (\bibinfo {year}
  {2006})}\BibitemShut {NoStop}%
\bibitem [{\citenamefont {Nakahara}(2003)}]{nakahara-2003}%
  \BibitemOpen
  \bibfield  {author} {\bibinfo {author} {\bibfnamefont {M.}~\bibnamefont
  {Nakahara}},\ }\href@noop {} {\emph {\bibinfo {title} {Geometry, Topology and
  Physics}}}\ (\bibinfo {year} {2003})\ \bibinfo {note} {{B}ristol, UK: Hilger
  (1990) 505 p. (Graduate student series in physics)}\BibitemShut {NoStop}%
\end{thebibliography}
\end{document}